\documentclass[12pt,usenames,dvipsnames]{article}

\usepackage{latexsym}
\usepackage{amssymb,amsfonts,amsmath}
\usepackage{graphicx} 
\usepackage{indentfirst}
\usepackage{bbm}
\usepackage{amssymb}
\usepackage{verbatim}
\usepackage{amsmath, amsthm,amssymb}
\usepackage{mathrsfs}
\usepackage{hyperref}
\usepackage{amsfonts}
\usepackage{dsfont}
\usepackage{cite}
\usepackage{xcolor}
\usepackage{enumerate}
\usepackage{cleveref}
\parskip=\medskipamount

\newcommand{\be}{\begin{equation}}
\newcommand{\ee}{\end{equation}}
\newcommand{\bea}{\begin{eqnarray}}
\newcommand{\eea}{\end{eqnarray}}
\newcommand{\non}{\nonumber}
\newcommand{\sU}{\mathsf{U}}

\numberwithin{equation}{section}

\begin{document}

\begin{titlepage}
\begin{flushright}
May, 2026 \\
Revised version: June, 2026
\end{flushright}
\vspace{5mm}

\begin{center}
{\Large \bf 
Causal self-dual nonlinear electrodynamics from 
\\
the Born-Infeld theory
}
\end{center}

\begin{center}

{\bf Sergei M. Kuzenko and Jonah Ruhl} \\
\vspace{5mm}

\footnotesize{
{\it Department of Physics M013, The University of Western Australia,\\
35 Stirling Highway, Perth W.A. 6009, Australia}}  
~\\
\vspace{2mm}
~\\
Email: \texttt{ 
sergei.kuzenko@uwa.edu.au, jonah.ruhl@uwa.edu.au}\\
\vspace{5mm}

\end{center}

\begin{abstract}
\baselineskip=14pt
Recently we have proposed a new auxiliary-field formulation for self-dual nonlinear electrodynamics (NLED) which makes use of two building blocks: (i) a seed self-dual theory  $L(F_{\mu\nu};g)$, where $F_{\mu \nu}$ is the electromagnetic field strength and $g$ a duality-invariant coupling constant; and (ii) a scalar potential $W(\psi)$. Our formulation is based on the Lagrangian $ \mathfrak{L}(F_{\mu\nu};\psi) = L(F_{\mu\nu};\psi) + W(\psi)$, where $\psi$ is an auxiliary scalar field. Integrating out $\psi$, using its equation of motion, one obtains a $\mathsf{U}(1)$ duality-invariant NLED. Different self-dual NLEDs are derived by choosing different potentials $W(\psi)$. In the case that the seed Lagrangian defines the Born-Infeld theory, in this paper we demonstrate that the resulting models for self-dual NLED are causal and provide a general solution of the self-duality equation. We also elaborate on the procedure to relate our formulation to that developed  by Russo and Townsend.
\end{abstract}
\vspace{10mm}

\vfill

\vfill
\end{titlepage}

\newpage

\section{Introduction} \label{section1}

In a recent paper \cite{kuzenko2026generalisations}, we proposed a new auxiliary-field formulation for $\sU(1)$ duality-invariant nonlinear electrodynamics (NLED) \cite{GZ1, BB, GR1,GR2,GZ2,GZ3}. It makes use of two building blocks:
(i)  a seed Lagrangian\footnote{Throughout this paper, the term `Lagrangian' stands for `Lagrangian density.'} 
$L(F_{\mu\nu};g)$, where $F_{\mu \nu}$ is the electromagnetic field strength and $g$ a coupling constant; and (ii) a scalar potential $W(\psi)$.  The seed Lagrangian describes 
a model for self-dual NLED in the sense that it solves the self-duality equation
\begin{align}
\label{S-DE}
    G^ {\mu \nu}\widetilde{G}_{\mu \nu}  +F^{\mu \nu}\widetilde{F}_{\mu \nu} =0~, \qquad 
        \widetilde{G}_{\mu\nu} :=\frac{1}{2}\varepsilon_{\mu\nu\sigma\rho}G^{\sigma\rho} = 2\,\frac{\partial {L}}{\partial F^{\mu\nu}}~.
\end{align}
This equation is the necessary and sufficient condition for the theory to possess  invariance under $\sU(1)$ duality rotations
which, in the infinitesimal case, look like
\bea
\delta F_{\mu \nu}  = \lambda G_{\mu \nu}  ~, \qquad \delta G_{\mu \nu}  = - \lambda F_{\mu \nu} ~.
\eea
 The coupling constant in $L(F_{\mu\nu};g)$ is a duality-invariant parameter. 
Replacing the parameter $g$ with a duality-invariant scalar field $\psi$ gives a self-dual theory $L(F_{\mu\nu};\psi)$.\footnote{This point was discussed long ago in the context of ${\cal N} = 1$ supersymmetric nonlinear electrodynamics \cite{KT1,KT2}.}
Adding the potential, $W(\psi)$, to $L(F_{\mu\nu};\psi)$ results in the Lagrangian proposed in
\cite{kuzenko2026generalisations},
\bea
\mathfrak{L}(F_{\mu\nu};\psi) = L(F_{\mu\nu};\psi) + W(\psi)~,
\label{potential}
\eea
which describes a self-dual theory in the sense that
$\mathfrak{L}(F_{\mu\nu};\psi)$ will solve the self-duality equation
\eqref{S-DE} for any choice of $W(\psi)$.

The scalar field $\psi$ is auxiliary since it appears in \eqref{potential} without derivatives. 
 It may be integrated out using its algebraic equation of motion 
 \bea
\frac{\partial}{\partial \psi}    \mathfrak{L}(F_{\mu\nu};\psi) =0~,
\label{EoM1.4}
\eea
 and then one arrives at a new self-dual theory theory $L(F_{\mu\nu}) =  \mathfrak{L}\big(F_{\mu \nu}; \psi (F)\big) $.\footnote{The Lagrangian  $L(F_{\mu\nu}) $ satisfies the self-duality equation \eqref{S-DE} since  $\frac{\partial {L}}{\partial F^{\mu\nu}} = \frac{\partial {\mathfrak L}}{\partial F^{\mu\nu}} + \frac{\partial {\mathfrak L}}{\partial \psi} \frac{\partial \psi} {\partial F^{\mu\nu}}
 = \frac{\partial {\mathfrak L}}{\partial F^{\mu\nu}} $.}
 Making different choices of $W(\psi)$ allows one to generate different models for self-dual NLED.
Choosing a different seed Lagrangian $L(F_{\mu\nu};g)$ leads to another auxiliary-field formulation for self-dual NLED. 

As a seed Lagrangian $L(F_{\mu\nu};g)$, one important choice is  the ModMax theory proposed by Bandos, Lechner, Sorokin and Townsend
\cite{bandos2020}
\begin{align}\label{ModMax}
    {L}_{\rm MM}(F_{\mu\nu};g)\equiv{ L}_{\text{MM}}(S,P; g)
      = S \cosh g  + \sqrt{S^{2}+P^{2}}  \sinh g~, \qquad g\geq 0~,
\end{align}
where 
\begin{align}\label{SP-invariants}
    S := -\frac{1}{4}F_{\mu\nu}F^{\mu\nu}~, \qquad
    P := -\frac{1}{4}F_{\mu\nu}\tilde{F}^{\mu\nu}~,
\end{align}
are the invariants of the electromagnetic field strength $F_{\mu\nu}$, and $g $ is the coupling constant.\footnote{The parameter $g$ was denoted $\gamma$ in \cite{bandos2020}.}
This choice of $L(F_{\mu\nu};g)$ leads to  the
Russo-Townsend auxiliary-field formulation \cite{russo2025simplified} of causal
self-dual NLED\footnote{The minus sign in front of the potential \eqref{ModMax2} corresponds to the definition ginev in \cite{russo2025simplified}.}
\begin{align}\label{ModMax2}
    \mathfrak{L}_{\rm MM}(S,P; \psi)
      = S \cosh \psi  + \sqrt{S^{2}+P^{2}}  \sinh \psi - {\cal W}(\psi)~.
\end{align}
Further studies of the Russo-Townsend formulation have appeared, e.g., in \cite{baglioni2026relatingauxiliary, Babaei-Aghbolagh:2025uoz, Babaei-Aghbolagh:2026vkm}.

In our work \cite{kuzenko2026generalisations},  another choice of seed Lagrangian $L(F_{\mu\nu};g)$ was explored,
\begin{align}\label{BI}
    {L}_{\text{BI}}(F_{\mu\nu}; g)\equiv{ L}_{\text{BI}}(S,P; g)=g-\sqrt{ g^{2} -2 g S-P^{2} }~, \qquad g>0~,
\end{align}
which describes the Born-Infeld (BI) theory \cite{Born:1934gh}.\footnote{The
coupling constant $g$ is often denoted $T$.} After promoting the duality-invariant
parameter to an auxiliary field, our formulation leads to\footnote{By freezing $\psi$ to a constant value $g>0$,  \eqref{BI-aux-field} reduces to the  BI theory, modulo an irrelevant additive constant $W(g)$. This feature is formal, since  $\psi$ is the auxiliary field which must be eliminated by making use of its algebraic equation. The equation has no solution $\psi = {\rm const}$.}
\begin{align}\label{BI-aux-field}
     \mathfrak{L}_{\rm BI}(S,P; \psi) = \psi - \sqrt{\psi^2 - 2\psi S - P^2} + W(\psi)~, \qquad \psi>0~.
\end{align}
The equation of motion for
$\psi$ is\footnote{One observes that the `vacuum field value $\psi_0$, which solves  the equation $W'(\psi_0)=0$, corresponds to $S=P=0$. 
%The solution $\psi_0$ is unique since $W'(\psi)$ is an increasing function due to \eqref{condition-potential}.
This property is similar to that present in the Russo-Townsend formulation \cite{russo2025simplified}. The uniqueness of $\psi_0$ follows from \eqref{condition-potential}.}
\begin{align}\label{BI-aux-field-eom}
 \frac{\partial}{\partial \psi}    \mathfrak{L}_{\rm BI}=   1 - \frac{\psi - S}{\sqrt{\psi^2 - 2\psi S - P^2}} + W'(\psi) = 0  \quad \implies
 \quad W'(\psi) \geq 0~. 
\end{align}
Since 
\bea
\Xi:= \frac{\partial^2}{\partial \psi^2}    \mathfrak{L}_{\rm BI}= \frac{S^{2}+ P^{2} }{(\psi^{2} -2\psi S - P^{2} )^{3/2} }+W''(\psi)
  ~,
 \label{ift}
 \eea
equation \eqref{BI-aux-field-eom}
admits a unique smooth solution,
$\psi = \psi(S,P)$ under the condition
\bea
W''(\psi) > 0 ~\implies ~ \Xi >0~.
\label{condition-potential}
\eea
Substituting this solution back into \eqref{BI-aux-field}
yields a self-dual theory ${L}(F_{\mu\nu})$. Different choices of the potential
$W(\psi)$ generate different models for self-dual NLED. For instance, the choice\footnote{The possible values of $\psi$ 
should be restricted to $0<\psi \leq g {\rm e}^{- \gamma}$. This restriction is obtained by requiring the following two conditions: (i) $W'(\psi) \geq 0$, in accordance with \eqref{BI-aux-field-eom}; 
and (ii) the argument of square root in \eqref{MMBpotential} should be non-negative. It also follows that 
$  W_{\text{MMB}}(\psi;g,\gamma =0)=0$.}
\begin{align}
    W_{\text{MMB}}(\psi;g,\gamma)=g-\psi-\sqrt{\psi^{2}-2g\psi \cosh \gamma + g^{2}}~,
    \label{MMBpotential}
\end{align}
with $g$ and $ \gamma $ positive parameters (of which $\gamma $ is  dimensionless), was shown in \cite{kuzenko2026generalisations} to produce, upon elimination of $\psi$, 
 the so-called `ModMaxBorn' theory \cite{bandos2020, Bandos:2020hgy}
\bea
\label{MMB}
    L_{\rm MMB}(S,P; g, \gamma)= g-\sqrt{g^{2}-2g L_{\rm MM}(S,P;\gamma) -P^{2}}~,
\eea
whilst a linear potential recovers the ModMax theory
\cite{kuzenko2026generalisations}.
In what follows, we will use notation $ \mathfrak{L}(S,P; \psi) $ for the Lagrangian \eqref{BI-aux-field}.

An important comment is required regarding the condition \eqref{condition-potential}.
In our previous paper   \cite{kuzenko2026generalisations},
%\cite{kuze order nko2026generalisations},
the weaker condition  $W''(\psi)\geq0$ was imposed. However, in the present paper we exclude  $W''(\psi)=0$ 
in order for the solution $\psi(S,P)$ to be analytic in a neighbourhood of 
$S=P=0$. For example, the potential corresponding to the ModMax theory 
\eqref{ModMax} is linear 
in $\psi$  \cite{kuzenko2026generalisations}, $W_{\rm MM} (\psi) = (\cosh  g -1) \psi$, and thus $W''_{\rm MM} (\psi) =0$. 
In this case, the solution\footnote{This solution is not defined in the Maxwell case, $g=0$.} 
$\psi = \psi(S,P)$ is 
\bea
\psi = S + \sqrt{S^2 +P^2} \, \coth  g~,
\eea
and it is not analytic at $S=P=0$.

Condition \eqref{condition-potential} guarantees that the Lagrangian of self-dual NLED, $L(S,P)$, which is obtained upon elimination of $\psi$, is smooth in $S$ and $P$. It is instructive to compare this condition with the one given long ago in \cite{KT2}.
Introducing the complex variable $\omega = -(S + {\rm i} P)$ and its conjugate $\bar \omega$, the Lagrangian $L(S,P)$ turns into
$L(\omega , \bar \omega)$. %If we represent 
Representing $L(\omega , \bar \omega)$ in the form
\bea
L(\omega , \bar \omega)  = -\frac 12 \, \Big( \omega + \bar{\omega} \Big) +
\omega \, \bar{\omega}  \Lambda (\omega, \bar{\omega} )~,
%\label{A.2}
\eea
 the self-duality equation \eqref{S-DE} becomes
\bea
{\rm Im} \bigg\{ \frac{\partial (\omega \, \Lambda) }{\partial  \omega}
- \bar{\omega}\,
\left( \frac{\partial (\omega \, \Lambda )  }{\partial \omega} \right)^2 \bigg\} = 0~,
\label{GZ4}
\eea
see \cite{KT2} for the technical details. Using this equation, one can show that $L(\omega , \bar \omega)$ is  analytic in a neighbourhood of $\omega =0$ provided $\Lambda (\omega , \bar \omega)$ is real analytic \cite{KT2}.

This letter is organised as follows.
In section \ref{section2} we demonstrate that our formulation 
\eqref{BI-aux-field} generates self-dual NLED theories satisfying strong causality conditions.
We also review the Courant-Hilbert solution for self-dual NLED \cite{GR1,GZ3} and discuss  the interplay between $\mathsf{U}(1)$ duality invariance and the causal structure of NLED theories.
Section \ref{section3} demonstrates the generality of our formulation.
Concluding comments are given in  section \ref{section4}.
Appendix \ref{appendixA} is devoted to the technical aspect of non-recoverability of seed theory.

%%%%%%%%%%%%%%%%%%%%%%%%%%%%%%%%%%
%%%%%%%%%%%%%%%%%%%%%%%%%%%%%%%%%

\section{Causality and self-duality}\label{section2}

In this section we demonstrate that our auxiliary-field formulation \eqref{BI-aux-field} generates causal self-dual NLED theories which provide a general solution to the self-duality equation \eqref{S-DE}. To start with, we discuss the relevant causality aspects.

\subsection{Causality in nonlinear electrodynamics}

Let us consider a model for NLED with Lagrangian $L(F_{\mu\nu}) =  L(S,P)$. Assuming the existence of a weak-field limit, 
the necessary and sufficient conditions for $L(S,P)$ to describe causal propagation were found by Schellstede, Perlick and 
L{\"ammerzahl \cite{Schellstede:2016zue}. They are: 
\begin{subequations}\label{CausCond}
\begin{align} 
L_S &>0~; \label{CausCond.a}\\
L_{SS}\geq 0~, \quad L_{PP}&\geq 0~, \quad L_{SS}L_{PP} - L_{SP}^2 \geq 0~; \label{CausCond.b}\\
L_S &> 2U L_{SS} +2VL_{PP} -2P L_{SP}~, \label{CausCond.c}
\end{align}
\end{subequations} 
where we have introduced the non-negative variables
\begin{align}\label{UV-invariants}
    U = \frac{1}{2}\big(\sqrt{S^2 + P^2} - S\big)~, \qquad
    V = \frac{1}{2}\big(\sqrt{S^2 + P^2} + S\big)~.
\end{align}
As usual,  the subscripts in \eqref{CausCond} denote partial derivatives, e.g. $L_S = \partial  L/\partial S$.
It was demonstrated in \cite{Bandos:2021rqy} that 
the relations   \eqref{CausCond.a} and  \eqref{CausCond.b} are also the conditions for strict convexity of $L$ as a function of the electric field $\vec E$.
Inequality \eqref{CausCond.c} takes a simpler form \cite{Russo:2024llm} 
\bea
L_U +2 U L_{UU} <0~,
\label{U-condition}
\eea
if the Lagrangian is considered as a functions of $U$ and $V$, $L(U,V)$.

It is well known that the Born-Infeld theory \eqref{BI} satisfies all the causality conditions \eqref{CausCond}.
Specifically, one finds
\begin{subequations}
\bea
L_{{\rm BI},S} = \frac{g}{\sqrt{g^{2} -2 g S-P^{2}}} &>&0~, \\
L_{{\rm BI},SS}= \frac{g^2}{ (g^{2} -2 g S-P^{2} )^{3/2} }&>&0~, \\
L_{{\rm BI},PP}= \frac{g^2-2gS}{ (g^{2} -2 g S-P^{2} )^{3/2} }&>&0~,\\
L_{{\rm BI},SS}L_{{\rm BI},PP} -(L_{{\rm BI},SP})^2 =\frac{g^2}{ (g^{2} -2 g S-P^{2} )^{2} }&>&0~.
\eea
\end{subequations}
Condition \eqref{U-condition}, which is equivalent to \eqref{CausCond.c},
 is also satisfied for the Born-Infeld theory, $L_{\rm BI} = g    - \sqrt{(g + 2U)(g - 2V)} $. One finds 
\bea
L_{{\rm BI},U} +2U L_{{\rm BI},UU} = -g \frac{(g-2V)^{1/2} }{(g+2U)^{3/2} }<0~.  
\eea

We are going to demonstrate that the NLED theory 
\bea
L(S,P) = \mathfrak{L} \big(S,P; \psi (S,P)\big)~, 
\eea 
with $\psi (S,P)$ the unique solution to the equation of motion \eqref{BI-aux-field-eom}, 
also satisfies all the causality conditions \eqref{CausCond}. Condition \eqref{CausCond.a} trivially holds
since 
\bea
L_S = \mathfrak{L}_S + \mathfrak{L}_\psi \psi_S = \mathfrak{L}_S = \frac{\psi}{ \sqrt{R}} >0~,
\label{BI-type-causality-S}
\eea
where we have introduced 
\bea
R =\psi^2 - 2\psi S - P^2>0~.
\eea
For second partial derivatives of $L$, one obtains the following expressions:
\bea
L_{SS} = \mathfrak{L}_{SS} - \Xi (\psi_S)^2~, \quad L_{PP} = \mathfrak{L}_{PP} - \Xi (\psi_P)^2~, \quad
L_{SP} = \mathfrak{L}_{SP} - \Xi \psi_S \psi_P~,
\eea
where $\Xi$ is defined by \eqref{ift}. By differentiating  the equation of motion \eqref{BI-aux-field-eom}, one also finds 
\bea
\psi_S= - \Xi^{-1} \mathfrak{L}_{S \psi} ~, \quad \psi_P= - \Xi^{-1} \mathfrak{L}_{P \psi} ~.
\eea
Now direct calculations give
\begin{subequations} \label{BI-type-causality}
\bea
L_{SS} &=& \Xi^{-1} \left\{  W'' (\psi ) \frac{ \psi^2  }{ R^{3/2} } +\frac{P^2 }{R^2} \right\} >0~,  \label{BI-type-causality.a}\\
L_{PP} &=& \Xi^{-1} \left\{  W'' (\psi ) \frac{ \psi^2 -2\psi S}{ R^{3/2} } +\frac{S^2 }{R^2} \right\} >0~, \label{BI-type-causality.b}\\
L_{SS}L_{PP} - (L_{SP})^2 &=&  \Xi^{-1}   W'' (\psi ) \frac{\psi^2}{R^{7/2} }>0~. \label{BI-type-causality.c}
\eea
\end{subequations}
Regarding the causality condition \eqref{U-condition}, we obtain 
\bea
L_U +2 U L_{UU} =- \Xi^{-1}  \left\{  W'' (\psi ) \frac{ \psi (\psi -2V)^2 }{ R^{3/2} } + 
\frac{(U+V)^2(\psi - 2V) }{R^2} \right\} <0~.
\label{BI-type-causality-U}
\eea
We conclude that all causality conditions \eqref{CausCond} are satisfied, and the strong version of \eqref{CausCond.b} holds.

It is of interest to see what happens to the above causality conditions in the case that
%limit
\bea
W''(\psi)=0\quad \implies \quad 
\Xi= \frac{S^{2}+ P^{2} }{(\psi^{2} -2\psi S - P^{2} )^{3/2} } = \frac{S^{2}+ P^{2} }{R^{3/2}}~.
\eea
Then, the condition \eqref{BI-type-causality-S} does not change, while the relations \eqref{BI-type-causality} turn into
\begin{subequations} 
\bea
L_{SS} = \frac{P^2 }{ \sqrt{R} (S^2+P^2) }  &\geq &0~,  
\\
L_{PP} = \frac{S^2 }{ \sqrt{R} (S^2+P^2) }  &\geq &0~,  
\\
L_{SS}L_{PP} - (L_{SP})^2 &=&  0~. 
\eea
\end{subequations}
Finally, the relation \eqref{BI-type-causality-U} takes the form
\bea
L_U +2 U L_{UU} = 
-\frac{(\psi - 2V) }{\sqrt{R} }  <0~.
\eea
We see that the causality conditions \eqref{CausCond} hold.

The above analysis of  causal self-dual NLED is based on the structure of the BI theory. It considerably differs from the Russo-Townsend analysis \cite{Russo:2024llm}, reviewed below in Section \ref{section2.3}, which made use of the Courant-Hilbert solution.

%%%%%%%%%%%%%%%%%%%%%%%%%%%%%%%%%%%%%%%%%%

\subsection{Courant-Hilbert solution}

Given a model for self-dual NLED, its Lagrangian $L(F_{\mu\nu})$ satisfies the self-duality equation \eqref{S-DE}.
Since  the  Lagrangian is a function of the invariants $S$ and $P$,  $L(F_{\mu\nu}) =  L(S,P)$, the self-duality equation 
can be written as
\begin{align}\label{self-duality-SP}
    P\!({ L}_S^2 - { L}_P^2 - 1)
    = 2S{ L}_S{ L}_P~.
\end{align}
This equation was derived for the first time by Bialynicki-Birula \cite{BB}.
In terms of the variables \eqref{UV-invariants},
equation \eqref{self-duality-SP} takes the following equivalent form \cite{GR1,GZ3}:
\begin{align}\label{self-duality-UV}
{ L}_{U}{ L}_{V}=-1~.
\end{align}
The general solution of this equation was given by several groups in the 1990s \cite{GR1,Perry:1996mk, GZ3} using the Courant-Hilbert method \cite{CourantH},
\begin{subequations}\label{CH-solution}
\begin{align} 
    L_{\rm CH} &= \ell(\tau) - \frac{2U}{\dot{\ell}(\tau)}~, \label{CH-solution.a}\\
    \tau &= V + \frac{U}{\dot{\ell}(\tau)^2}\geq 0~, \label{CH-solution.b}
\end{align}
\end{subequations}
where $\ell(\tau)$ is a function of $ \tau $ such that $\dot \ell \neq 0$. Once $\ell(\tau)$ has been chosen, relation \eqref{CH-solution.b}
allows one to express $\tau$ as a function of $U$ and $V$. Then, plugging the obtained function $\tau (U,V)$ into \eqref{CH-solution.a}
results in a Lagrangian $L(U,V)$ which solves the self-duality equation \eqref{self-duality-UV}. Indeed, one may check that 
\bea
L_U = - {\dot \ell}^{-1}, \qquad L_V = \dot \ell~,
\eea
see \cite{GR1,GZ3, Perry:1996mk, Russo:2024llm} for the derivation.

It should be pointed out that the variables \eqref{UV-invariants} are suitable to describe those models for NLED which are parity even, 
which means $L(S,-P) = L(S,P)$. All models for self-dual NLED with a weak-field limit,  are parity even
(see \cite{KT2} for the technical details).
%%%%%%%%%%%%%%%%%%%%%%%%%%%%%%%%%%%%

\subsection{Self-duality and causality} \label{section2.3}

In the case of self-dual NLED, the causality conditions \eqref{CausCond} can be recast in terms of the function $\ell(\tau)$ 
which determines the Courant-Hilbert solution \eqref{CH-solution}. This has been done by Russo and Townsend \cite{Russo:2024llm}. 
First of all, the causality condition \eqref{CausCond.a} proves to be  equivalent to 
\bea
\dot{\ell} > 0~. \label{precondition}
\eea
Secondly, imposing the convexity conditions \eqref{CausCond.b} 
and \eqref{CausCond.c}  leads to 
\bea
\dot{\ell} \geq 1~, \qquad \ddot{\ell} \geq 0~.
\label{CH-caus1}
\eea
The second relation shows that $\ell (\tau)$ is a convex function.

Let us now assume $\ell (\tau)$ to satisfy only the condition \eqref{precondition} and to be strictly convex, 
\bea
\ddot{\ell} > 0~.
\label{CH-caus2}
\eea
Then  a Legendre transform $\Omega( y)$ of $\ell(\tau)$ exists. 
Following \cite{russo2024dualities}, it is obtained by defining a new variable $y = \dot{\ell}$ and introducing
\bea
\Omega(y) = \Big( y \tau - \ell(\tau) \Big)\Big|_{\tau = \tau(y)} \quad \implies \quad \tau = \Omega'(y)~.
\label{CH-Legendre}
\eea
The variable $y$ was 
interpreted in \cite{russo2025simplified}
as an auxiliary scalar field, and $\Omega(y)$ as its scalar
potential. 
Plugging the relations 
\bea
\ell(\tau) =   \tau y  -\Omega(y)~, 
\qquad
\dot{\ell} = y~,\qquad  \tau = \Omega'(y)~,
\eea
in the Courant-Hilbert solution \eqref{CH-solution} leads to the Lagrangian
\begin{align}
 L_{\text{RT}}(U,V;y) &= yV - \frac{U}{y} - \Omega(y)~,
\qquad y > 0~. 
\label{RT-formulation}
\end{align}
The equation of motion for $y$, $\partial L_{\rm RT}/\partial y = 0$, is
\begin{align}
\Omega'(y) &= V + \frac{U}{y^2}~, \label{RT-eom}
\end{align}
which is equivalent to \eqref{CH-solution.b}. 
The causality conditions \eqref{CH-caus1} and \eqref{CH-caus2} turn into 
\cite{russo2025simplified}
\begin{align}
    y &\geq 1~, \qquad \Omega''(y) > 0~. \label{RT-causality}
\end{align}
 
It was observed in  \cite{russo2025simplified} that 
the ModMax-type auxiliary-field formulation \eqref{ModMax2} follows from \eqref{RT-formulation}
by defining 
$y := {\rm e}^{\psi}$ and $ \Omega(y):= {\cal W}(\psi)$. 
The causality conditions \eqref{RT-causality} turn into
\bea
\psi \geq 0~,  \qquad {\cal W}''(\psi) > {\cal W}'(\psi)~.
\eea

In a recent publication \cite{baglioni2026relatingauxiliary}, 
it has been shown that the auxiliary-field model \eqref{RT-formulation}
is related to the so-called $\mu$-frame version of the Ivanov-Zupnik formulation for self-dual NLED theories
\cite{IZ_N3, IZ1, IZ2}.

%%%%%%%%%%%%%%%%%%%%%%%%%%%%%%%%%%%%
%%%%%%%%%%%%%%%%%%%%%%%%%%%%%%%%%%%%

\section{Generality} \label{section3}

It remains to show how to relate the model \eqref{RT-formulation} to our formulation  \eqref{BI-aux-field}. 
Written in terms of $U$ and $ V $,
\eqref{BI-aux-field} takes the form
\begin{align}\label{BI-UV}
     L(U,V;\psi) &= \psi
    - \sqrt{(\psi + 2U)(\psi - 2V)} + W(\psi)~.
\end{align}
The equation of motion for $\psi$, $\partial L/\partial \psi = 0$, is
\begin{align}
    1 - \frac{\psi + U - V}{\sqrt{(\psi + 2U)(\psi - 2V)}}
    + W'(\psi) &= 0. \label{eom-UV}
\end{align}

We require the Lagrangians  \eqref{RT-formulation} and \eqref{BI-aux-field} to coincide provided the corresponding auxiliary fields are integrated out using the equations of motion \eqref{RT-eom} and \eqref{eom-UV}, respectively. That is,
\bea
L_{\text{RT}} \big(U,V; y(U,V)\big) =  L\big(U,V;\psi(U,V)\big)~.
\eea 
This gives 
\bea
   y = \sqrt{\frac{\psi + 2U}{\psi - 2V}}~.
\eea 
Plugging this relation into  \eqref{RT-formulation} leads to a relation between the potentials 
\bea
\Omega(y) &=& V  \sqrt{\frac{\psi + 2U}{\psi - 2V}} - U \sqrt{\frac{\psi - 2V}{\psi + 2U}}
+  \sqrt{(\psi + 2U)(\psi - 2V)} -\psi - W(\psi) \non \\
&=&  \psi W'(\psi) - W(\psi)~.
\label{Omega}
\eea
It may be shown that 
\begin{align}
    W'(\psi) &= \frac{y}{2} + \frac{1}{2y} - 1~.
    \label{Wprime-y}
\end{align}
It is assumed above that $y $ and $\psi $ are functions of $U$ and $V$. The relations \eqref{Omega} and \eqref{Wprime-y}
allow one, in principle,  
to reconstruct $\Omega(y)$ from the potential $W(\psi)$.

Our analysis confirms that the BI-type formulation \eqref{BI-UV} is suitable to describe general  causal self-dual
NLEDs with the exception of the seed BI theory (see the appendix for the technical details).

%%%%%%%%%%%%%%%%%%%%%%%%%%%%%%%%%%%%%%%

\section{Concluding comments} \label{section4}

In this paper we elaborated on the auxiliary-field formulation for self-dual NLED \eqref{potential} proposed in 
\cite{kuzenko2026generalisations}. Choosing the BI Lagrangian as seed Lagrangian, we demonstrated that our model 
\eqref{BI-aux-field} generates causal self-dual NLEDs with a weak-field limit under the condition $W''(\psi)>0$
on the scalar potential. Choosing a weaker condition on the potential, $W''(\psi) \geq 0$, also leads to causal self-dual NLED theories including the ModMax theory. We also showed how our formulation is related to that proposed earlier by Russo and Townsend
\cite{russo2025simplified}. 

In addition to the auxiliary-field formulations for self-dual NLED developed in \cite{kuzenko2026generalisations, russo2025simplified},
and the more general Ivanov-Zupnik approach\footnote{The Ivanov-Zupnik formulation 
is truly universal in the sense that it has been extended to: (i) ${\cal N} =1$ and ${\cal N}=2$ supersymmetric models for self-dual nonlinear electrodynamics \cite{K13,ILZ}; (ii) self-dual theories in $4n$ dimensions \cite{Kuzenko:2019nlm};  (iii) self-dual models for $\cal N$-extended superconformal gauge multiplets \cite{KR21-2, Kuzenko:2023ebe}; and (iv) higher-derivative deformations of self-dual NLED models, including the ModMax theory \cite{Kuzenko:2024zra}.} 
\cite{IZ_N3, IZ1, IZ2}, there exist other auxiliary-field formulations for self-dual NLED 
\cite{Hatsuda:1999ys, BN, PST2, INZ, Buratti:2019cbm, Avetisyan:2022zza}.
Perhaps the most important virtue of our formulation \eqref{BI-aux-field} is that it is based on the Born-Infeld theory which plays a fundamental role in string theory  \cite{Fradkin:1985qd, Leigh:1989jq} and in the framework of models for partial ${\cal N}=2 \to {\cal N}=1$ supersymmetry breaking on maximally supersymmetric four-dimensional backgrounds \cite{BG,RT,KT-M16}, as well as in ${\cal N}=2$ super Yang-Mills theories with partially broken and nonlinearly realised supersymmetry \cite{Farakos:2025oru}.
Our formulation \eqref{BI-aux-field} is expected to be more useful for quantum-mechanical applications than \eqref{ModMax2}
since, unlike  \eqref{ModMax2}, our Lagrangian is analytic in the field strength. We also hope to apply our formulation for constructing new models for self-dual NLED; this will be discussed elsewhere.

Another interesting feature of our formulation \eqref{potential} is that 
its equation of motion \eqref{EoM1.4} 
%corresponding to our model \eqref{potential} 
has an  interpretation in terms of the energy-momentum tensor. 
Since the parameter $g$ in $L(F_{\mu\nu};g)$ is duality invariant, it is well known that $\partial L(F_{\mu\nu};g)/\partial g$ is a duality-invariant observable \cite{GZ2,GZ3}. It is also known that this observable may be expressed in terms of the energy-momentum tensor\footnote{This theorem extends  several explicit examples considered earlier in the literature in the context of $T\bar T$ deformations \cite{Conti:2018jho, Babaei-Aghbolagh:2022uij, Ferko:2022iru}.} 
\cite{Ferko:2023wyi},
\bea
\frac{\partial}{\partial  g}  L(F_{\mu\nu};g) = {\mathfrak F}(T_{\mu\nu}; g)~.
\eea
Thus the equation of motion \eqref{EoM1.4}  can be recast in the form 
\bea
{\mathfrak F}(T_{\mu\nu}; \psi) + W'(\psi) =0~.
\eea
This equation tells us that the dynamics of $\psi$ is determined by the energy-momentum tensor. 

There are two simple procedures to generate one-parameter families of self-dual NLED from a given self-dual model theory
$L(F) =  L(S,P)$.
\begin{itemize} 
\item The rescaling map \cite{KT2}
\bea
L(F) ~\to ~L^{ (\gamma ) } (F) := \frac{1}{\gamma^2} L(\gamma F)~, \qquad \gamma \in {\mathbb R}^+ ~.
\eea
Its application to the  ModMax \eqref{ModMax} and Born-Infeld \eqref{BI} theories gives
\bea
 L^{(\gamma)}_{\rm MM}(F;g) =  {L}_{\rm MM}(F;g)~, \qquad
 L^{(\gamma)}_{\text{BI}}(F; g) =  {L}_{\text{BI}}(F; g\gamma^{-2})~.
\eea
\item The ModMax map\footnote{This construction generalises earlier perturbative results derived in the context of 
$T\bar T$-flows \cite{Babaei-Aghbolagh:2024uqp}.} 
 \cite{Murcia:2025psi}
\bea
L(S,P)~\to ~ L^{[\gamma]} (S,P) := L\Big({L}_{\rm MM}(S,P; \gamma), P\Big)~.
\eea 
Its application to the  ModMax \eqref{ModMax} and Born-Infeld \eqref{BI} theories gives
\bea
 L^{[\gamma]}_{\rm MM}(F;g) =  {L}_{\rm MM}(F;g+\gamma)~, \qquad
 L^{[\gamma]}_{\text{BI}}(F; g) =  {L}_{\text{MMB}}(F; g, \gamma)~,
\eea
with ${L}_{\text{MMB}}(F; g, \gamma)$ being the ModMaxBorn theory \eqref{MMB}.
\end{itemize}
It would be interesting to explore ${L}_{\text{MMB}}(F; g, \gamma)$ in the role of seed Lagrangian for \eqref{potential}.

\noindent
{\bf Acknowledgements:}\\
The work of SMK is supported in part by the Australian Research Council, project DP230101629.

\appendix

\section{Non-recoverability of seed theory} \label{appendixA}

In \cite{kuzenko2026generalisations} it was shown that  particular choices of the 
 potential $W(\psi)$ in \eqref{BI-aux-field} lead to the ModMaxBorn and ModMax theories upon integrating out the auxiliary scalar $\psi$.
For the BI-type auxiliary field model \eqref{BI-aux-field}, it is possible to show that
there exists no potential $ W(\psi) $ such that one can recover the Born-Infeld Lagrangian \eqref{BI}
after integrating out the auxiliary field $ \psi. $  

\textbf{Proof.} 
Let us assume that there \textit{does} exist some potential $
W_{\text{BI}}(\psi) $, and define 
\begin{align}\label{Lh}
    \hat{ L}(S,P;\psi):=\psi-\sqrt{\psi^{2} -2\psi S -P^{2}}+
W_{\text{BI}}(\psi)~,
\end{align}
where it is understood that $ \psi=\psi(S,P)$ is a solution to 
\begin{align}\label{BI-proof-eom}
    1-\frac{\psi-S}{\sqrt{\psi^{2} -2\psi S -P^{2}}} +W_{\text{BI}}'(\psi)=0~.
\end{align}
For a solution $ \psi(S,P) $ to exist we require  $ W_{\text{BI}}''(\psi)\geq
0 $ then under this assumption we should find that \eqref{BI} and \eqref{Lh} are
equivalent, which would imply equality of partial derivatives
\begin{align}\label{Lds}
    \frac{\partial{ L_{\text{BI}}}}{\partial S}=\frac{g}{\sqrt{g^{2}-2gS-P^{2}}}~, \qquad 
    \frac{\partial\hat{ L}}{\partial S}=\frac{\psi}{\sqrt{\psi^{2}-2\psi S-P^{2}}}~.
\end{align}
Equating the results in \eqref{Lds} and rearranging for $ \psi $ yields
\begin{align}
    \psi=g ~~ \text{or} ~~ \psi=\frac{-P^{2} g }{P^{2}+2gS}~.
\end{align}
In the first case we find that
$\hat{ L}(S,P)= L_{\text{BI}}(S,P;g)+W_{\text{BI}}(g)$ and hence require
$W_{\text{BI}}(g)=0$. Evaluating \eqref{BI-proof-eom} at $S=P=0$ then yields
$W_{\text{BI}}'(g)=0$, from which the general equation of motion becomes,
\begin{align}
        1-\frac{g-S}{\sqrt{g^{2} -2g S -P^{2}}}=0\implies S^{2}+P^{2}=0~,
\end{align}
which contradicts the relations \eqref{ift} and \eqref{condition-potential}.
In the second case, we note that $\psi<0$ which contradicts the
condition in \eqref{BI-aux-field}. This completes the proof.
%$\square$ 

Our conclusion above  is analogous to that in \cite{russo2025simplified}, in which
the seed ModMax theory was not explicitly recoverable.

\begin{footnotesize}

\end{footnotesize}

\end{document}